\pdfoutput=1
\documentclass[
reprint,
superscriptaddress,
preprintnumbers,
nofootinbib,
 amsmath,amssymb,
 aps,
prd,
]{revtex4-2}

\usepackage{CJKutf8}
\usepackage[T5]{fontenc}
\usepackage[utf8]{inputenc}
\usepackage[table,svgnames,dvipsnames]{xcolor}
\definecolor{Blue}{HTML}{1c1e94}
\usepackage[colorlinks=true,citecolor=Blue,linkcolor=Blue,urlcolor=Blue, backref=false,pdfborder={0 0 0}]{hyperref}
\usepackage[normalem]{ulem}
\usepackage{aas_macros}
\usepackage{graphicx}% Include figure files
\usepackage{dcolumn}% Align table columns on decimal point
\usepackage{bm}% bold math
\usepackage[mathlines]{lineno}% Enable numbering of text and display math
\usepackage{orcidlink}
\usepackage{cases}% for dealing with mathematics,
\usepackage{booktabs}
\usepackage{comment}
\usepackage{multirow}
\usepackage{makecell}
\usepackage{siunitx}
\usepackage{booktabs}
\usepackage{xcolor}
\graphicspath{{figs/}}

\usepackage{fontawesome5}
\makeatletter
\newcommand{\github}[1]{%
  \href{#1}{\faGithub}%
}
\makeatother

\newcommand\numberthis{\addtocounter{equation}{1}\tag{\theequation}}

\newcommand{\code}[1]{{\texttt{#1}}}
\newcommand{\desilike}{\code{desilike}}

\newcommand{\base}{\texttt{c000}}
\newcommand{\lowsigma}{\texttt{c004}}
\newcommand{\abacus}{\textsc{AbacusSummit}}
\newcommand{\abacuspng}{\textsc{AbacusPNG}}
\newcommand{\abacushod}{\textsc{AbacusHOD}}
\renewcommand{\emph}[1]{\textit{#1}}

\usepackage[nameinlink,noabbrev]{cleveref}
\crefname{equation}{Eq.}{Eqs.}
\crefname{section}{Sec.}{Secs.}
\crefname{figure}{Fig.}{Figs.}
\crefname{table}{Tab.}{Tabs.}
\crefname{appendix}{App.}{Apps.}
\Crefname{figure}{Figure}{Figures}
\Crefname{equation}{Equation}{Equations}
\Crefname{section}{Section}{Sections}
\Crefname{table}{Table}{Tables}
\Crefname{appendix}{Appendix}{Appendices}

\newcommand{\<}{\left\langle}
\renewcommand{\>}{\right\rangle}

\newcommand{\const}{\rm const.}
\newcommand{\bphi}{b_\phi}
\newcommand{\fnl}{f_{\rm NL}}

\renewcommand{\L}{\mathcal{L}}
\newcommand{\M}{\mathcal{M}}

\renewcommand{\max}{\mathrm{max}}

\newcommand{\kmax}{k_\max}

\newcommand{\degsq}{\,{\rm deg}^{2}}

\newcommand{\Om}{\Omega_m}

\newcommand{\rhom}{\rho_{\rm m}}
\newcommand{\rhobm}{\bar{\rho}_{\rm m}}

\newcommand{\nbar}{\bar{n}}
\newcommand{\dm}{\delta_{\mathrm{m}}}

\newcommand{\Pshot}{P_{\rm shot}}

\newcommand{\deltacr}{\delta_{\rm cr}}

\newcommand{\Mpchcubed}{\, h^{-3} \, \rm Mpc^{3}}
\newcommand{\Gpchcubed}{\, h^{-3} \, \rm Gpc^{3}}
\newcommand{\Msunh}{\,h^{-1} M_\odot}

\newcommand{\hmpcinv}{\,h\,{\rm Mpc^{-1}}}

\newcommand{\vrr}{\bm{r}}
\newcommand{\vx}{\bm{x}}

\newcommand{\vk}{\bm{k}}

\definecolor{RoyalBlue}{rgb}{0.25,.41,.88}
\definecolor{WildStrawberry}{HTML}{EE2967}
\newcommand{\rep}[1]{#1}

\begin{document}
\begin{CJK*}{UTF8}{gbsn}
\title{How I stop worrying about non-universality and $\bphi$:\\Constraining local $\fnl$ with $\bphi$ priors from HOD posteriors}

\author{Jiaxi Yu (\CJKfamily{gbsn}禹佳希)\orcidlink{https://orcid.org/0009-0001-7217-8006}}
\email{jiaxi.yu@ipmu.jp}
\affiliation{Kavli IPMU (WPI), UTIAS, The University of Tokyo, 5-1-5 Kashiwanoha, Kashiwa, Chiba 277-8583, Japan}
\affiliation{Center for Data-Driven Discovery, Kavli IPMU (WPI), UTIAS, The University of Tokyo, Kashiwa, Chiba 277-8583, Japan}
\author{Nhat-Minh Nguyen (Nguyễn Nhật Minh)\orcidlink{0000-0002-2542-7233}}
\email{nhat.minh.nguyen@ipmu.jp}
\affiliation{Kavli IPMU (WPI), UTIAS, The University of Tokyo, 5-1-5 Kashiwanoha, Kashiwa, Chiba 277-8583, Japan}
\affiliation{Center for Data-Driven Discovery, Kavli IPMU (WPI), UTIAS, The University of Tokyo, Kashiwa, Chiba 277-8583, Japan}
\affiliation{Institute For Interdisciplinary Research in Science and Education, ICISE, Quy Nhon, 55121, Vietnam}
\date{\today}

\begin{abstract}
Local primordial non-Gaussianity (local PNG) induces a scale-dependent
contribution to galaxy clustering proportional to $\fnl\,\bphi$, where
$\fnl$ is the local PNG amplitude and $\bphi$ encodes the galaxy response
to a long-wavelength primordial potential perturbation. Uncertainty in
$\bphi$ is the dominant obstacle to precise, robust constraints on $\fnl$
from galaxy surveys. We translate small-scale clustering constraints on
the galaxy--halo connection into priors on $\bphi$: sampling the posterior
of a halo occupation distribution (HOD) model fit to the DESI EDR, we
generate mocks from which we measure $\bphi$ and construct its prior.
Validating against additional mocks with different local PNG amplitudes,
we show that the method recovers consistent true $\fnl$ values, even in the presence of
assembly bias. \rep{Code and data products} are available at
\href{https://github.com/MinhMPA/bphi-prior-from-hod-posterior}{\faGithub\ MinhMPA/bphi-prior-from-hod-posterior}.
\end{abstract}

\preprint{IPMU26-0027}

\maketitle
\end{CJK*}

\section{Introduction}
\label{sec:introduction}

The near-Gaussianity of the primordial density field is a precision null test of the simplest models of inflation. Any significant departure from Gaussianity would have to be a fossil of additional degree of freedoms and interactions active during inflation. The strongest constraints on primordial non-Gaussianities (PNG) today still come from the cosmic microwave background anisotropies measured by the Planck experiment \cite{Planck:2019kim}, but constraints from galaxy clustering probed by galaxy redshift surveys are approaching, and will eventually surpass that level of precision (see, e.g., Figs.~6-8 of Ref.~\citep{Chen:2026src}).

\rep{For galaxy clustering and large-scale structure, PNG of the local type (local PNG) produces a characteristic, approximately $k^{-2}$ bias contribution through primordial coupling between long and short modes \cite{Dalal:2007cu}. Other sources of scale dependence include massive neutrinos, modified gravity and relativistic/projection effects. We distinguish these contributions and state our fiducial cosmological assumptions in \cref{sec:scale-dependent_bias}.}
Recent analyses from the Data Release 1\footnote{\url{https://data.desi.lbl.gov/doc/releases/dr1/}} (DR1, \cite{DESI_DR1}) of the Dark Energy Spectroscopic Instrument \citep[e.g.,][]{Chaussidon:2024qni,Rosado-Marin:2026kpx,Brown:2026cul} have focused on this channel and demonstrated that DESI data can already deliver competitive constraints on local PNG compared to Planck.

This promise comes with a catch. In galaxy clustering, the local PNG corrections appear as products of $\bphi\fnl$, in which $\fnl$ describes the amplitude of the local PNG signal while $\bphi$ encapsulates the response of galaxy formation to the PNG signal.
Without robust knowledge of $\bphi$, galaxy clustering data constrains only the product $\bphi\fnl$, and neither the $\fnl$ central value of nor the uncertainty can be reliably inferred from the measurements \cite{Barreira:2020ekm,Barreira:2022sey, Lazeyras:2022koc}.

A common shortcut is to assume the universality relation, which fixes $\bphi$ from the (Gaussian) linear bias $b_1$ under the assumption of a universal halo mass function. As discussed in \cref{sec:universality_relation}, this holds for idealized, mass-selected tracers such as simulated halos, but is a poor approximation for the color- and flux-selected galaxies that surveys actually target \citep[e.g.,][]{Barreira:2020kvh,Barreira:2021fuf}. Robust $\fnl$ constraints therefore require prior information on $\bphi$, and two routes have been proposed to obtain it.

The first route attempts to calibrate $\bphi$ against cosmological simulations, optionally anchored to external observations. Because $\bphi$ carries a secondary dependence on tracer properties beyond $b_1$, most importantly halo concentration and assembly history, this approach models that dependence explicitly.
Ref.~\citep{Sullivan:2023qjr} trained a machine-learning model on the observable properties of simulated \textsc{IllustrisTNG} galaxies to predict $\bphi$ for samples that mimic the DESI emission-line galaxies (ELGs) and luminous red galaxies (LRGs).
Ref.~\citep{Hadzhiyska:2025rez} related $\bphi$ to the higher-derivative Lagrangian bias $b_{\nabla^2\delta}$, an effective concentration proxy internal to the galaxy bias expansion, calibrated on \abacus mocks of DESI LRGs and quasars.
Ref.~\citep{Moore:2026glz} instead marginalized a semi-analytic model of galaxy formation over external observables, specifically the stellar mass function and the stellar-to-halo mass relation, to construct a $\bphi$ prior conditioned on those observables. In all of these, the $\bphi$--tracer connection is calibrated to simulations, and any observational input enters through generic galaxy-formation observables rather than through the clustering of the target sample itself.

The second route suggests to infer $\bphi$ directly from the survey data. Refs.~\citep{Dalal:2025eve,Sullivan:2025fie} estimated the local-PNG response of the tracer abundance from the observed redshift evolution of its comoving number density, $dn/dz$, modulo a residual term encoding departures from universality and selection effects. This approach is attractive because it uses only directly observable quantities, but its practical utility hinges on reliably modeling the residual contribution induced by survey selection effects.

In this work, we propose a third route that combines simulations with the observed small-scale clustering of the target-selected samples. For each sample, we use the galaxy--halo connection as described by the halo occupation distribution (HOD) model to map the observed small-scale clustering to a prior distribution of $\bphi$ in the local PNG analysis. As a demonstration, we sample the HOD posteriors constrained by the DESI LRG data, from the ``One-Percent Survey'' (SV3) in the DESI Early Data Release\footnote{\url{https://data.desi.lbl.gov/doc/releases/edr/}} (EDR), reported by Ref.~\citep{Yuan:2023ezi}, to generate pairs of HOD mocks (at a standard and a low $\sigma_8$), then measure $b_\phi$ using the separate universe technique for all sampled mock pairs to construct the $\bphi$ priors.

The rest of this paper is organized as follows. In \cref{sec:scale-dependent_bias} and \cref{sec:universality_relation}, we briefly review the scale-dependent bias signature of local PNG and the universality relation between $b_1$ and $\bphi$, plus departure thereof. We continue to describe the separate-universe method and measurement of $\bphi$ in \cref{sec:bphi_seperate-universe} and \cref{sec:simulation}. In \cref{sec:inference} and \cref{sec:results}, we describe the inference framework and discuss the results, before concluding in \cref{sec:conclusion}.

\section{Local PNG--scale-dependent bias}
\label{sec:scale-dependent_bias}

The primordial gravitational (Bardeen) potential $\phi$, in the presence of local PNG, can be parametrized as
\begin{equation}
\phi(\vx)=\phi_G(\vx)+\fnl\left[\phi_G(\vx)^2-\<\phi^2_G(\vx)\>\right]
\label{eq:PNG_Bardeen}
\end{equation}
where $\phi_G$ is a Gaussian random field and $\fnl$ sets the amplitude of the quadratic, non-Gaussian term, which vanishes in the standard single-field scenario.

In the slow-roll limit, single-field inflation models obey the consistency relation that essentially predicts $\fnl\simeq\frac{5}{12}(n_s-1)$ \cite{Maldacena:2002vr}, where $n_s$ is the spectral index of the power spectrum of primordial density fluctuations at the end of inflation, with the scale-invariant primordial power spectrum corresponds to $n_s=1$.

Current constraints on $n_s$ from cosmic microwave background experiments \cite{Planck:2018vyg,AtacamaCosmologyTelescope:2025blo} imply that a detection of $|\fnl|\gtrsim\mathcal{O}(1)$ would violate the consistency relation hence rule out the simplest single-field inflation picture. Reaching the precision of $\sigma_{\fnl}~\mathcal{O}(1)$ in constraints of $\fnl$ is therefore one of the key science goals for current surveys such as DESI~\citep{DESI:2016fyo} and SPHEREx~\citep{SPHEREx:2014bgr}, and for next-generation spectroscopic surveys like MUST~\citep{Zhao:2024alp,cai2026}, Spec-S5~\citep{Spec-S5:2025uom}, and WST~\citep{WST:2024rai}.

\rep{The Bardeen potential in \cref{eq:PNG_Bardeen} is related to the linear matter density fluctuations at redshift $z$, $\dm(\vx,z)\equiv\frac{\rhom(\vx,z)}{\rhobm(z)}-1$, by}
\begin{equation}
\dm(\vk,z)=\M(k,z)\phi(\vk),
  \label{eq:Bardeen_deltam}
\end{equation}
\rep{where $\M(k,z)=\frac{2}{3}\frac{k^2T(k)D(z)}{\Om H_0}$, with $T(k)$ and $D(z)$ being the matter transfer function and the linear growth factor, respectively. At leading order in $\fnl$, Eqs.~\eqref{eq:PNG_Bardeen} and \eqref{eq:Bardeen_deltam} give the linearly evolved primordial local-PNG contribution to the matter bispectrum,}
\rep{
\begin{align*}
B_{\rm mmm}^{\rm prim}(k_1,k_2,k_3)&=\M(k_1)\M(k_2)\M(k_3)B_{\phi\phi\phi}(\vk_1,\vk_2,\vk_3)\\
&=\M(k_1)\M(k_2)\M(k_3)\times\\
&\left[2\fnl\left(P_{\phi\phi}(k_1)P_{\phi\phi}(k_2) + 2\,\text{perm.}\right)\right]\numberthis
\label{eq:PNG_bispectrum}
\end{align*}
}
\rep{Nonlinear gravitational evolution generates an additional contribution even for Gaussian initial conditions.}

\cref{eq:PNG_bispectrum} peaks in the squeezed limit, where one long-wavelength mode $\vk_1$ couples to two much shorter modes $(\vk_2,\vk_3)$. In configuration space, local PNG therefore modulates the small-scale variance from one patch to another. For biased tracers, this modulation appears as a large-scale correction to the power spectrum,
\begin{equation}
P_{\rm gg}(k,z)=\left[b_1(z)+\frac{b_\phi(z)\,\fnl}{\M(k,z)}\right]^2 P_{\rm mm}(k,z) + \Pshot(k,z)
\label{eq:PNG_Pgg}
\end{equation}
where the first term inside the square bracket describes galaxy clustering with a Gaussian Bardeen potential and the Gaussian bias coefficient $b_1$, while the second term describes the modulation by local PNG with the local PNG bias coefficient $\bphi$. The final term, $\Pshot$, is the shot-noise contribution.

\rep{On sufficiently large scales, the local-PNG bias correction in \cref{eq:PNG_Pgg} scales as $k^{-2}$. This response to primordial coupling between long and short modes is absent from ordinary local gravitational bias for Gaussian adiabatic initial conditions~\citep{Assassi:2015png}. It is convenient to absorb this term into an effective bias,}
\begin{equation}
\tilde{b}_1(k,z)
\equiv
b_1(z)
+
\frac{b_\phi(z) \fnl}{\M(k,z)} .
\label{eq:eff bias}
\end{equation}
\rep{Here $\M(k,z)$ is computed in the fixed fiducial late-time $\Lambda$CDM cosmology specified in \cref{sec:simulation}, with standard gravity. }
\rep{The galaxy power spectrum in real-space and redshift-space are then written as follows, respectively \citep{Kaiser:1987qv,Hamilton1992},}
\begin{align}
P_{\rm gg}(k,z)
&=
\tilde{b}_1(k,z)^2P_{\rm mm}(k,z)
+
\Pshot(k,z),
\label{eq:real space pk}
\\
P^S_{\rm gg}(k,\mu,z)
&=
\left[
\tilde{b}_1(k,z)
+
f(z)\mu^2
\right]^2
P_{\rm mm}(k,z)
\nonumber\\
&+
\Pshot(k,z),
\label{eq:redshift space pk}
\end{align}
where $\mu$ is the cosine of the angle between the Fourier mode and the line of sight, and $f\approx\Omega_m^{0.55}(z)$ is the linear growth rate. The monopole and quadrupole are
\begin{align}
P^{S,\,\ell=0}_{\rm gg}(k,z)
&=
\left[
\left(\tilde{b}_1(k,z)\right)^2
+
\frac{2}{3} \tilde{b}_1(k,z) f(z)
+
\frac{1}{5} f^2(z)
\right]
\nonumber\\
&\qquad\times
P_{\rm mm}(k,z)
+
\Pshot(k,z)\label{eq:PSgg_ell0}.
\\
P^{S,\,\ell=2}_{\rm gg}(k,z)
&=
\left[
\frac{4}{3} \tilde{b}_1(k,z) f(z)
+
\frac{4}{7} f^2(z)
\right]
P_{\rm mm}(k,z)\label{eq:PSgg_ell2}.
\end{align}
\rep{Our mock analysis uses a Newtonian description of redshift-space clustering. Massive neutrinos can generate a bias transition near the free-streaming scale~\citep{LoVerde:2014bias}, while modified gravity can produce scale-dependent growth and bias~\citep{Parfrey:2011bias}. Relativistic/projection effects in observed number counts, which are omitted in our Newtonian mocks, can also contain $k^{-2}$-like contributions with coefficients that depend on redshift and sample selection~\citep{Jeong:2012gr}. These contributions must be assessed when interpreting an observed large-scale modulation as primordial.}

\section{Bias---universality relation}
\label{sec:universality_relation}

The Gaussian and local PNG bias $b_1$ and $\bphi$ characterize the responses of the abundance of the biased tracers to local variations of the matter background density $\rhobm$ and the matter fluctuation amplitude $\sigma_8$, respectively. Specifically,
\begin{align}
  b_1(M,z) &= \frac{d\ln n(M,z)}{d\ln\rhobm(z)}\label{eq:b1_su_response},\\
  b_\phi(z) &= 2\frac{d\ln n(M,z)}{d\ln\sigma_8(z)}\label{eq:bphi_su_response}
\end{align}
The second relation follows from the fact that, as noted above, local PNG modulates the amplitude of small-scale fluctuations within each local patch.

If the tracer abundance $n(M,z)$ is universal, i.e., if it can be written as a function $f(\nu)$ of only the peak height $\nu=\deltacr^2/\sigma^2(M,z)$, where $\deltacr$ is the threshold overdensity for spherical collapse linearly extrapolated to $z=0$ and $\sigma^2(M,z)$ is the variance of matter fluctuations within the Lagrangian patch corresponding to the enclosed mass $M$, then \cref{eq:b1_su_response} and \cref{eq:bphi_su_response} are related by the so-called ``universality relation''\footnote{Formally, \cref{eq:universality_relation} also requires the additional assumption of tracer conservation such that $b^{\rm Eulerian}_1=b^{\rm Lagrangian}_1+1$.}:
\begin{equation}
b_\phi(M,z)=2\deltacr\left[b_1(M,z)-1\right].\label{eq:universality_relation}
\end{equation}
Most observational constraints on $\fnl$ from galaxy data fit for $b_1$ and treat $\bphi$ as a derived quantity through this relation. Strictly speaking, however, \cref{eq:universality_relation} is expected to hold only for idealized tracers selected by virial mass, such as dark matter halos in N-body simulations. \rep{It has been proved, \citep[by e.g., ][]{Slosar2008,Reid2010,Barreira:2020ekm,Barreira:2022sey} that it is a poor approximation for observed tracers selected by properties like the formation history, stellar mass, star formation rate, or color-magnitude cuts.}

In this work, as outlined in \cref{sec:introduction}, we measure the possible $\bphi$ range (i.e., the $\bphi$ prior) from the separate-universe response, using pairs of galaxy mocks with high and low $\sigma_8$. We describe the measurement in details in the next section.

\section{$\bphi$ as a separate-universe response}
\label{sec:bphi_seperate-universe}

The key insight behind the separate-universe approach is that, in the presence of local PNG, a long-wavelength mode of primordial potential fluctuations rescales the amplitude of the power spectrum of matter fluctuations on smaller scales. Keeping all other cosmological parameters fixed, the $\bphi$ response can be written in terms of $\sigma_8$ as
\begin{equation}
  b_{\phi}=2\frac{\partial\ln \bar n}{\partial\ln\sigma_8}.
  \label{eq:bphi_su_response_partial}
\end{equation}
%This is the same response measured in Ref.~\citep{Barreira:2021fuf}, but expressed through $\sigma_8$ rather than $A_s$; see also Ref.~\citep{Dalal:2025eve,Sullivan:2025fie} for the same normalization.
\cref{eq:bphi_su_response_partial} can be estimated using finite difference around a fiducial cosmology, for example, as
\begin{equation}
  \widehat b_{\phi}
  =
  2\,
  \frac{\ln N-\ln N^{\rm fid}}
     {\ln\sigma_8-\ln\sigma_8^{\rm fid}} ,
  \label{eq:bphi_su_response_fd}
\end{equation}
where $N$ is the number of galaxies, and the superscript $i$ denotes a cosmology that differs from the fiducial one only in $\sigma_8$. In this work, we use pairs of galaxy mocks generated from $\abacus$ $\base$ and $\lowsigma$ simulations which differ only by their $\sigma_8$ values (see \cref{sec:AbacusSummit_base_lowsigma8} for details).

\section{$\bphi$ from separate-universe mocks}
\label{sec:simulation}

\subsection{$\abacus$ simulations}
\label{sec:AbacusSummit}

We use the $\abacus$ simulation suite\footnote{\url{https://abacussummit.readthedocs.io/en/latest/simulations.html}}~\citep{Maksimova:2021ynf} to generate mock galaxy samples. The suite was run with \textsc{Abacus}~\citep{Garrison:2021lfa}, an $N$-body gravity calculation that splits the gravitational force into direct near-field summation and a far-field multipole convolution. This split allows large-volume simulations to retain accurate forces down to the halo scales relevant for this work. The public $\abacus$ release includes simulations at the fiducial cosmology, and other secondary cosmologies.

Halos in $\abacus$ are identified with the \textsc{CompaSO} halo finder\footnote{\url{https://abacussummit.readthedocs.io/en/latest/compaso.html}}~\citep{Hadzhiyska:2021zbd}. \textsc{CompaSO} is an on-the-fly spherical-overdensity halo finder designed for \textsc{Abacus} simulations. It uses competitive particle assignment to deblend neighboring
halos and to assign particles near halo boundaries. These choices enter our analysis through the halo masses, centers, and secondary halo properties that the HOD model later takes as input.

\subsubsection{$\abacus$ base and low-$\sigma_8$ sets}
\label{sec:AbacusSummit_base_lowsigma8}

Given a posterior of HOD parameters constrained by small-scale clustering statistics, we estimate $\bphi$ using \cref{eq:bphi_su_response_fd}, with pairs of HOD mocks constructed from the $\abacus$ base simulation\footnote{Details of cosmological parameters are available on \url{https://abacussummit.readthedocs.io/en/latest/cosmologies.html}} (\base; the Planck18 $\Lambda$CDM cosmology in Ref.~\citep{Planck:2018vyg}, with $\sigma_8=0.807952$) and its low-$\sigma_8$ counterpart (\lowsigma; same cosmology as \base~except for $\sigma_8=0.749999$). Each simulation contains $N_{\rm part}=6912^3$ particles in a $V_{\rm sim}=2^3\Gpchcubed$ physical comoving volume, providing a mass resolution of $M_{\rm part}=2.1\times 10^9\, \Msunh$.
In order to construct samples closely mimicking DESI EDR LRG samples, we use the $z=0.5,\,0.8,\,1.1$ snapshots. There are six realizations of $\base$-$\lowsigma$ simulation pairs. \rep{The $\bphi$ variance from the cosmic variance derived from these realizations is either much smaller than the variations obtained from galaxy--halo relation, or does not influence the final $\fnl$ results.} So we ignore the influence of the cosmic variance hereafter and only use the first realization \texttt{ph000} to populate the mock galaxies.

\subsubsection{$\abacuspng$}
\label{sec:AbacusSummit_png}

The validation of the $\bphi$ priors requires another set of galaxy mocks generated from the $\abacuspng$ simulation set~\citep{Hadzhiyska:2024kmt}. The $\abacuspng$ simulations have the same box size as the $\abacus$ simulations used for the prior construction, $L_{\rm box}=2\,h^{-1}{\rm Gpc}$, but a lower particle mass resolution of $1.01\times10^{10}\Msunh$. Although coarser than the prior-construction simulations, this resolution remains sufficient for DESI LRGs. For these tests, we use the first realization, \texttt{ph000}, at the $z=0.5,0.8,1.1$ snapshots, considering the baseline cosmology, $\base$, and the $\fnl=+30$ and $\fnl=-30$ simulations, labeled \textsc{c300} and \textsc{c301}, respectively.

\subsection{$\abacushod$ Mocks}
\label{sec:desi_hod}

We generate galaxy mocks using the $\abacushod$\footnote{\url{https://github.com/abacusorg/abacusutils}} framework \cite{Yuan:2021izi}, which was developed for the $\abacus$ simulations. Its standard vanilla HOD model \cite{Zheng:2007zg} parameterizes the halo occupation number $\langle N\rangle$ of central galaxies and satellite galaxies as a function of halo mass $M$ as follows:
\begin{align}
  \langle N_{\rm cen}(M)\rangle
  &= \frac{f_{\rm ic}}{2}\,
  {\rm erfc}\left[
  \frac{\log_{10}(M_{\rm cut}/M)}{\sqrt{2}\sigma}
  \right],
  \label{eq:hod_ncen}\\
  \langle N_{\rm sat}(M)\rangle
  &= \langle N_{\rm cen}(M)\rangle
  \left(\frac{M-\kappa M_{\rm cut}}{M_1}\right)^\alpha ,
  \label{eq:hod_nsat}
\end{align}
Here $\sigma$ tunes the steepness of the error function from a constant central-galaxy occupation $f_{\rm ic}$ to 0. $f_{\rm ic}$ is the completeness of the galaxy sample, and thus not necessarily 1. $\kappa M_{\rm cut}$ specifies the minimum halo mass to host a satellite galaxy, and $M_1$ represents the typical host-halo mass of satellites. $\alpha$ is the power-law index of the satellite number. We apply $\abacushod$ that was calibrated by DESI EDR LRG samples at $0.4<z<0.6$, $0.6<z<0.8$~\citep{Yuan:2023ezi} on the simulation snapshots at redshifts $z=0.5, 0.8$ to construct our LRG1 and LRG2 mocks, respectively. For the $z=1.1$ snapshot, we adopt the best-fit parameters of the DESI LRGs at $0.95<z<1.1$ instead of those from $0.8<z<0.95$ for a more realistic high-redshift LRG HOD, given the redshift evolution of DESI LRGs at $0.8<z<1.1$.

\rep{To measure $\bphi$ and construct the $\bphi$ priors, we generate the galaxy mock pairs using the same HOD. We choose to inflate the width of the HOD posteriors to ensure that this HOD is both realistic and also inclusive of other surveys. The broadened HOD ensemble is constructed by drawing complete parameter vectors from the DESI HOD posteriors, according to their posterior weights, and replacing the parameter $\theta$ by ${\theta}_{s}=\langle {\theta} \rangle+s\times({\theta}-\langle {\theta} \rangle)$, where $\langle {\theta} \rangle$ is the posterior-weighted mean of that parameter, and $s$ is the rescale factor. This transformation scales the standard deviations of HOD parameters by $s$ while preserving the correlation coefficients and the shape of the joint parameter distribution. We choose $s=3$ for LRG1 and LRG2, and $s=1$ for LRG3 due to its weak constraints on HOD parameters. When sampling the HOD parameters, we reject non-finite values or violations of the physical bounds $\sigma>0$ and $\alpha,\kappa\geq0$, and require more than $10^5$ galaxies in each mock constructed from the $\base-\lowsigma$ simulation pair.}

\begin{figure*}[!t]
    \centering
    \includegraphics[width=\linewidth]{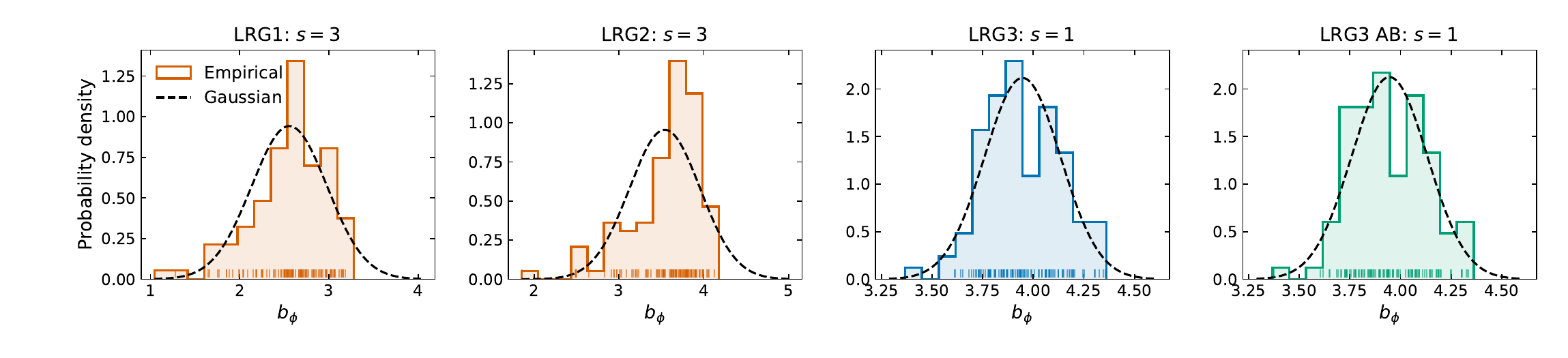}
    \caption{\rep{The $\bphi$ distributions for LRG1, LRG2, LRG3 and LRG3 with assembly bias, each constructed from 100 pairs of galaxy mocks. We use the broadened $s=3$ ensembles for LRG1 and LRG2 and the original $s=1$ ensemble for both LRG3 cases. The assembly-bias case uses $B_{\rm cen}=-0.04$ and $B_{\rm sat}=-0.17$. The normalized histograms and short vertical marks show the empirical responses; dashed curves are Gaussians with the same mean and variance as the 100 responses.}}
    \label{fig:bphi priors}
\end{figure*}

\rep{In the end, there are 100 accepted HOD parameter sets, and thus 100 galaxy mock pairs for each LRG sample. We compute the corresponding $\bphi$ values\footnote{The 100 $\bphi$ values for all redshift bins are provided in \url{https://github.com/MinhMPA/bphi-prior-from-hod-posterior/blob/master/data/bphi_prior_samples.csv}. } using \cref{eq:bphi_su_response_fd}. \Cref{fig:bphi priors} shows the empirical distributions and Gaussians matched to their means and variances, and the values are listed in \cref{tab:bphi priors}. The fiducial inference uses the empirical values. We test the effect of the Gaussian approximation of $\bphi$ distributions on \(\fnl\) separately in \cref{app:gaussian-bphi}.}

\begin{table}[ht]
\centering
\caption{\rep{The mean and variance of LRG $\bphi$ sampled from the DESI EDR HOD parameters. LRG3 and LRG3$_{\rm AB}$ represent priors constructed from the HOD implementations without and with assembly bias, respectively.}}
\begin{tabular}{ccccc}
\hline
& LRG1 & LRG2 & LRG3 & LRG3$_{\rm AB}$\\
\hline
 mean & 2.56 & 3.54 & 3.95 & 3.95 \\
 variance & $0.42^2$ & $0.42^2$ & $0.19^2$ & $0.19^2$ \\
\hline
\label{tab:bphi priors}
\end{tabular}
\end{table}

We then prepare the HOD mocks for validation of the constructed $\bphi$ priors. These mocks are generated from $\abacuspng$ simulations with the central value of the best-fit HOD parameters listed in Tables 3-4 of Ref.\citep{Yuan:2023ezi}. These best-fit HOD parameter sets are only implemented once to the first realization $\texttt{ph000}$ for $\fnl=0,\pm 30$.

\subsection{Assembly bias}
The assembly histories of galaxies and their host halos can play a role in determining the $\bphi$ value in observations \cite{Lazeyras:2022koc,Fondi:2023egm,Fondi:2026ilz}. To quantify potential impacts of assembly bias on the $\bphi$ priors and the final $\fnl$ constraints, we consider the HOD model with assembly bias \citep{Hearin:2015jnf} in which the galaxy occupation number depends on halo secondary properties at fixed mass. We implement the halo assembly bias measured by Ref.~\cite{Yuan:2021izi} for the LRGs from the Sloan Digital Sky Survey (SDSS), i.e., the halo-environment assembly bias \{$B_{\rm cen}=-0.04,\,B_{\rm sat}=-0.17$\}). In these HOD mocks with assembly bias, at a fixed halo mass, halos in high-density environments tend to host more galaxies while halos in low-density environments tend to host fewer galaxies, compared to the fiducial mocks without assembly bias. Therefore, the amplitude of the galaxy clustering with assembly bias is different from that of the fiducial mocks. \rep{The mean and variance, as well as the distribution of $\bphi$ obtained by HOD with assembly bias, are presented in the last column of \cref{tab:bphi priors} and \cref{fig:bphi priors}, respectively. Both show that adding SDSS-LRG-type assembly bias leads to negligible change in the $\bphi$ distribution. }

% For our test, we introduce assembly bias to both the HOD mocks from which we measure and construct the $\bphi$ priors (i.e. the mocks based on the $\abacus$ simulations) as well as the HOD mocks which we use as data vectors (i.e. the mocks based on the $\abacuspng$ simulations).
%\mn{I think this is redundant and confusing now that we have updated the description in \cref{sec:assembly bias}}

\section{Inference}
\label{sec:inference}

\subsection{Data Vectors and Covariances}
\label{sec:data vec and cov}

For each sample, we consider the redshift-space power spectrum monopole and quadrupole $\ell=0,2$ of the HOD galaxies as the observables. We measure the power spectrum multipoles in the $\abacuspng$ galaxy mocks using \texttt{pypower}\footnote{\url{https://github.com/cosmodesi/pypower}}
which uses the FFT-based estimator \cite{feldman_power_1994, hand_optimal_2017}
\begin{align}
  &P_{\rm gg}^{S, \ell}(k) =
  (2\ell + 1)
  \int \frac{d\Omega_k}{4\pi}
  \nonumber\\
  &\left[
  \int d\vrr_1
  \int d\vrr_2
  F(\vrr_1)F(\vrr_2)
  e^{i\mathbf{k}\cdot(\vrr_1-\vrr_2)}
  \mathcal{L}_{\ell}(\hat{\mathbf{k}}\cdot\hat{\vrr}_h)
  -
  P_{\rm noise}^{\ell}
  \right],
\end{align}
where \(\vrr_h=(\vrr_1+\vrr_2)/2\) and the Poisson noise is assumed to be zero, except for the monopole $\ell=0$ where
\begin{equation}
P_{\rm noise}^{\ell=0}=(1+\alpha)\int d\vrr\nbar(\vrr)w^2(\vrr)\mathcal{L}_{\ell}(\hat{\mathbf{k}}\cdot\hat{\vrr}_h).
    \label{eq:Poisson_noise}
\end{equation}

For a general survey,
\begin{align}
  F(\vrr) =
  \frac{
  w(\vrr)
  \left[
  n_g(\vrr)-\alpha n_s(\vrr)
  \right]
  }
  {A_g^{1/2}},
  \nonumber\\
  A_g =
  \int d\vrr\,
  \left[
  \bar{n}(\vrr)w(\vrr)
  \right]^2 ,
\end{align}
with \(\alpha=N_g/N_s\).
For periodic boxes, the selection is uniform,
so the random-catalog correction reduces to the usual mean-density subtraction
and \cref{eq:Poisson_noise} reduces to the usual Poisson shot noise $1/\nbar$.
The box size is \(L=2000\,h^{-1}{\rm Mpc}\).
We assume the line-of-sight direction to be the \(z\)-axis and assign particles to the mesh with TSC interpolation. We rebin the \texttt{pypower} outputs into $dk=0.002$. 
\rep{For the fitting range, we use \(k_{\rm min}=0.006\hmpcinv\) to keep as much large-scale information as possible. we also examine the reponses of $\fnl$ constraints to $\kmax$ in \cref{app:kmax_robustness}. Finally, our fiducial cuts are \(\kmax=(0.09,0.09,0.10)\hmpcinv\) for LRG1, LRG2 and LRG3, respectively. The smaller $\kmax$ for LRG1 and LRG2 is because of the non-linear gravitational evolution at low redshifts.}

For the covariance matrix, given the $k$-range of the data vector (see above) and the periodic-box geometry of the simulations we consider in this paper, we can neglect the trispectrum and super-sample contributions to the covariance. Therefore, the data-vector covariance is diagonal in $k$ (see, e.g., Eq.~(57) of Ref.~\citep{Wadekar:2019rdu}). We compute the covariance analytically using \texttt{thecov}\footnote{\url{https://github.com/cosmodesi/thecov}}\cite{Alves2026prep} using the power spectrum multipoles of the $\abacuspng$ mocks with $\fnl=0$ as the fiducials.

While implementing a periodic-box geometry, we scale the covariance matrices to match the corresponding volumes of the DESI samples. Specifically, we use the hypothetical effective volume $V_{\rm eff}$ of DESI LRGs in \cref{tab:veff} instead of the simulation volume $8\Gpchcubed$, computed as
\begin{equation}
  V_{\rm eff} = \sum_i \left(\frac{\bar{n}(z_i)P_0}{1+\bar{n}(z_i)P_0}\right)^2 \Delta V(z_i),
  \label{eq:Veff}
\end{equation}
where $\Delta V(z_i)$ is the comoving survey volume and $\overline{n}(z_i)$ is the mean number density of the tracer inside the redshift bin $z_i$, provided by DESI DR1~\cite{DESI_DR1,DESI2024II}, and $P_0=10,000\Mpchcubed$. We choose $17,000\degsq$ for LRG1 and LRG2 from DESI Year-8 sky coverage, but $14,000\degsq$ for LRG3 from DESI Year-5 coverage as a conservative estimation. 

\begin{table}[ht]
\centering
\caption{The hypothetical sky coverage and the effective volume of the complete DESI LRG samples.}
\begin{tabular}{lccc}
\hline
 & LRG1 & LRG2 & LRG3 \\
\hline
$A\degsq$ & $17,000$ & $17,000$ & $14,000$ \\
$V_{\rm eff}\Gpchcubed$ & $2.86$ & $4.42$ & $5.16$ \\
\hline
\label{tab:veff}
\end{tabular}
\end{table}

\subsection{Power spectrum model with $\fnl$ }
We use \desilike\footnote{\url{https://desilike.readthedocs.io/}} to build the local-PNG model power spectrum model with a scale-dependent bias \cref{eq:eff bias}. Its redshift-space power spectra \cref{eq:redshift space pk} is parameterized as
\begin{align}
P^S_{\rm gg}(k,\mu)
&=
J_{\rm AP}\,
F_{\rm FoG}(k,\mu;\Sigma_s)
\nonumber\\
&\times\left[
\tilde{b}_1(k,z)
+ f(z)\mu^2
\right]^2
P_{\rm mm}(k,z)
+ sn_{0},
\end{align}
where $J_{\rm AP}$ is the Alcock-Paczynski Jacobian factor, $F_{\rm FoG}(k,\mu;\Sigma_s)$ is the Fingers-of-God damping factor that suppresses redshift-space power along the line of sight due to small-scale random velocities, $\tilde{b}_1$ is the effective bias (\cref{eq:eff bias}), and $ sn_{0}$ is the shot noise term. Therefore, the nuisance parameter set in redshift bin $i$ is \{$b_{1,i}, \Sigma_{s,i}, sn_{0,i}$\}. 

The different parameterizations of the scale-dependent bias lead to different modes in $\desilike$. Without an external prior on $b_{\phi,i}$, the meaningful constraints are those on $B_i$, defined as 
\begin{equation}
  B_i \equiv b_{\phi,i}\fnl ,
  \label{eq:bfnl_definition}
\end{equation}
where the index $i$ labels a tracer at a redshift bin. This is the \texttt{bfnl} mode and the effective bias becomes
\begin{equation}
b_{\rm eff}(k,z_i) = b_{1,i} + B_i\,\alpha(k,z_i).
\end{equation}
When introducing our $\bphi$ priors, we would be able to activate the \texttt{bphi} mode, which treat $\bphi$ and $\fnl$ as separate parameters like in \cref{eq:eff bias}. Finally, if we assume the universality relation, 
\begin{equation}
b_{\phi,i} = 2\delta_c\left(b_{1,i}-p_i\right), \label{eq:fixed_p_prior}
\qquad \delta_c = 1.686,
\end{equation}
the pipeline is in \texttt{b-p} mode, where 
\begin{equation}
b_{\rm eff}(k,z_i)
=
b_{1,i}
+
2\delta_c\left(b_{1,i}-p_i\right)
f_{\rm NL}^{\rm loc}
\alpha(k,z_i).
\end{equation}
In our study, we fix \(p_i=1\).

\subsection{Priors}
\label{sec:prior}
%\mn{This needs better rephrasing to clearly separate the cases. A more consistent choice would be to always refer to them as ``weak prior'' and ``HOD-posterior prior'' (or ``HOD-informed'' or ``HOD-derived'' prior) throughout.}

\rep{In the \texttt{bfnl} analysis, the sampled PNG parameter in each redshift bin is \(B_i=b_{\phi,i}\fnl\), with a uniform prior \(B_i\in[-3000,3000]\). A posterior for \(\fnl\) is not determined by \(B_i\) alone: for this baseline case, we adopt uniform, independent \(b_{\phi,i}\in[-10,10]\) and a uniform \(\fnl\in[-300,300]\) when reconstructing its marginal posterior. We will refer to the baseline as ``weak prior'' hereafter. }
%This broad, symmetric-to-zero priors of $\bphi$ are assumptions, and the resulting \(\fnl\) posterior is sensitive to its support.}

\rep{For the analysis with informative $\bphi$ priors, we sample the common $\fnl\in[-300,300]$ directly in \texttt{bphi} mode. In each redshift bin, we use the empirical $\bphi$ prior shown in \cref{fig:bphi priors}, sampling from these 100 HOD-derived $\bphi$ values. The $\bphi$ priors for different redshift bins are assumed to be independent. This case will be referred to as ``empirical prior''. }

%The latent response is marginalized along with the other model parameters. We separately compare these results with moment-matched Gaussian response priors and, for Gaussian priors, with a product-amplitude parameterization in \cref{app:gaussian-bphi}. The latter comparison does not identify an empirical prior with a Gaussian one.

For the \texttt{b-p} comparison, \(\fnl\) is sampled directly with the same flat range and \(p=1\) is fixed in the universality relation. The remaining nuisance parameters \{$b_{1}, \Sigma_{s}, sn_{0}$\}, use the default \desilike\ flat priors. 

\subsection{Likelihood}
\label{sec:likelihood}

For each galaxy and redshift bin, the data vector is the joint redshift-space galaxy power-spectrum multipole vector
\begin{equation}
  \bm{d}_i
  =
  \left\{P^{S,\ell=0}_{\rm gg}(k_a,z_i),
      P^{S,\ell=2}_{\rm gg}(k_a,z_i)\right\}_{a=1}^{N_k}.
  \label{eq:data_vector_pk02}
\end{equation}
The monopole and quadrupole are fit together using the analytical covariance matrix $C_i$ from \code{thecov} matching to the clustering of $\abacuspng$ $\base$ mock. The likelihood for a given galaxy and redshift bin is
\begin{align*}
  -2\ln \L_i
  =&
  \left[\bm{d}_i-\bm{m}_i(\theta_i,\fnl)\right]^{T}
  C_i^{-1}
  \left[\bm{d}_i-\bm{m}_i(\theta_i,\fnl)\right]
  \\&+
  \const ,
  \numberthis\label{eq:gaussian_pk_likelihood}
\end{align*}
where $\bm{m}_i$ is the model vector evaluated for the same multipoles, $k$ bins, and covariance ordering as the data. %We use the \desilike\ PNG model, with nuisance parameters including the Gaussian bias and residual shot noise. The PNG contribution enters through either the parameter product $B_i$ or through $b_{\phi,i}\fnl$, depending on the prior choice described above.
The joint likelihood for multiple snapshots is the product of the independent bin likelihoods,
\begin{equation}
  \L(\fnl,\Theta)
  =
  \prod_i \L_i ,
  \label{eq:joint_likelihood}
\end{equation}
where $\Theta$ denotes all tracer-dependent nuisance parameters. We use \textsc{emcee} to obtain posterior distributions. In \texttt{b-p} and \texttt{bphi} modes, \(\fnl\) is sampled directly; for the weak \texttt{bfnl} comparison, it is reconstructed as described in \cref{app:margpost}.

\section{Results}
\label{sec:results}

\subsection{Validation}
\label{sec:validation}
\begin{figure*}[!t]
  \centering
  \includegraphics[width=\linewidth]{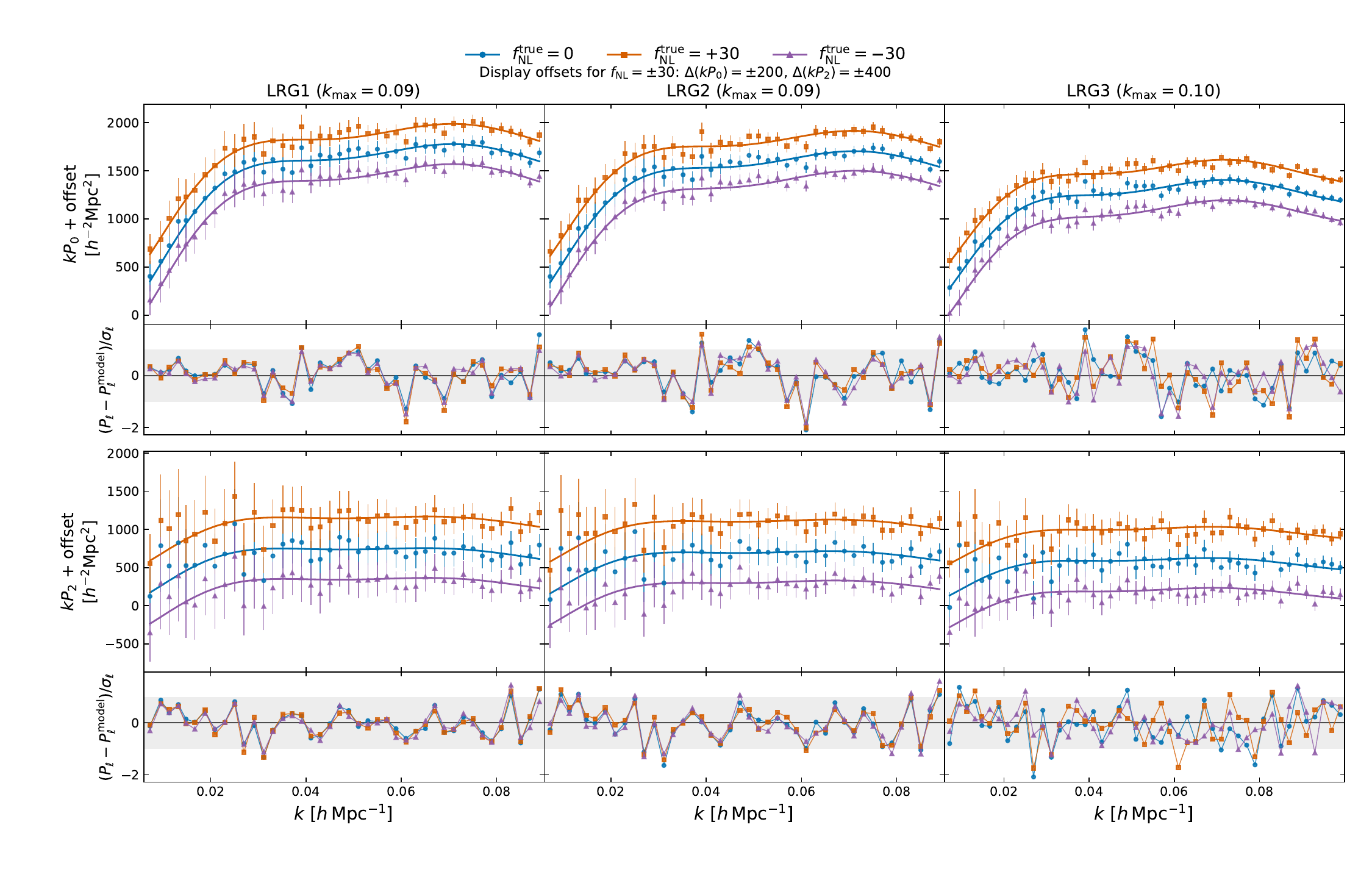}
  \caption{\rep{The power spectra monopole (upper pair of rows) and quadrupole (lower pair) for LRG1, LRG2 and LRG3 single-tracer fittings. Points and error bars show the mock data and diagonal standard deviations of the analytic covariance; lines show the best-fit theory of each single-tracer \texttt{bphi}-mode fit with its empirical \(b_\phi\) prior. Different colours and symbols denote the three PNG values of the simulations. For clarity, \(kP_0\) is shifted by \(\pm200\) and \(kP_2\) by \(\pm400\) for \(\fnl^{\rm true}=\pm30\), in \(h^{-2}{\rm Mpc}^{2}\); residuals \((P_\ell-P_\ell^{\rm model})/\sigma_\ell\) are unshifted, and the shaded bands indicate \(\pm1\). The displayed ranges use the fiducial tracer-dependent \(\kmax\).}}
  \label{fig:pk-fits}
\end{figure*}

\begin{figure*}[!t]
  \centering
  \includegraphics[width=0.85\linewidth]{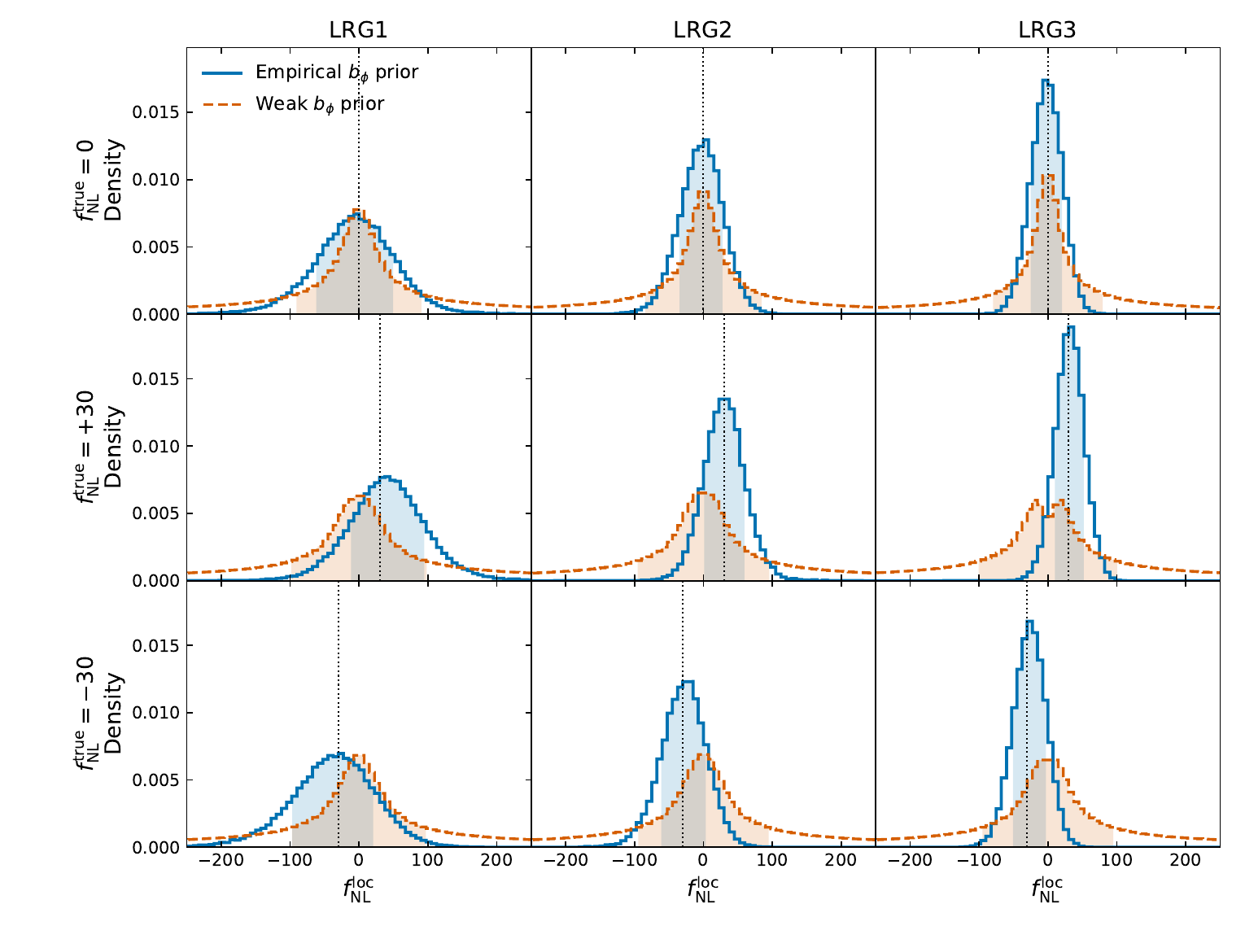}
  \caption{\rep{Marginalized \(\fnl\) posteriors for the three snapshots (columns) and input amplitudes \(0,+30,-30\) (rows). Blue solid curves show direct \texttt{bphi}-mode fits with empirical HOD priors; orange dashed curves reconstruct the \texttt{bfnl} weak-prior chains using \(b_\phi\in[-10,10]\). Vertical dotted lines mark the input values. Curves are smoothed for display; the intervals in \cref{tab:bphi-prior-snapshot} use the retained samples or the unsmoothed reconstruction.}}
  \label{fig:posterior consistency}
\end{figure*}

\rep{We validate the empirical $\bphi$ priors constructed using the separate-universe method with galaxy mocks having local PNG amplitudes of $\fnl^{\rm true}=0,+30,-30$. The best-fitting models in \cref{fig:pk-fits} reproduce the power-spectrum data vectors. \Cref{tab:bphi-prior-snapshot} presents the posterior medians and central 68\% credible intervals, while \cref{fig:posterior consistency} shows the $\fnl$ posteriors. The columns correspond to LRG1, LRG2, and LRG3, and the rows correspond to
$\fnl^{\rm true}=0,+30,-30$. For the weak prior baseline, \(\fnl\) is reconstructed from the \texttt{bfnl} chains using the broad \(b_\phi\) and \(\fnl\) priors stated in \cref{sec:prior,app:margpost}. In the informative empirical $\bphi$ prior cases, $\fnl$ is sampled directly in \texttt{bphi} mode. }

\begin{table}[tb]
\centering
\caption{\rep{Snapshot-by-snapshot marginalized \(\fnl\) constraints. The weak rows use \texttt{bfnl} product chains and the broad reconstruction prior, while the empirical rows use direct \texttt{bphi} fits. The reported improvement is \(100(1-W_{68}^{\rm empirical}/W_{68}^{\rm weak})\%\), where \(W_{68}\) is the full central 68\% interval width.}}
\label{tab:bphi-prior-snapshot}
\begin{tabular}{ccccc}
\hline
\(\fnl^{\rm true}\) & & LRG1 & LRG2 & LRG3 \\
\hline
\multirow{3}{*}{\(0\)}
& weak prior & \(0.0^{+90.7}_{-90.7}\) & \(0.0^{+84.5}_{-84.5}\) & \(0.0^{+79.7}_{-79.7}\) \\
& empirical prior & \(-5.2^{+54.9}_{-56.9}\) & \(-2.4^{+30.2}_{-32.1}\) & \(-1.6^{+22.3}_{-23.3}\) \\
& Improv. & 38.3\% & 63.2\% & 71.4\% \\
\hline
\multirow{3}{*}{\(+30\)}
& weak prior & \(0.0^{+98.3}_{-98.3}\) & \(0.0^{+95.5}_{-95.5}\) & \(0.0^{+100.7}_{-100.7}\) \\
& empirical prior & \(40.9^{+53.7}_{-52.6}\) & \(30.2^{+29.5}_{-29.2}\) & \(31.7^{+20.5}_{-21.4}\) \\
& Improv. & 45.9\% & 69.3\% & 79.2\% \\
\hline
\multirow{3}{*}{\(-30\)}
& weak prior & \(0.0^{+97.1}_{-97.1}\) & \(0.0^{+94.7}_{-94.7}\) & \(0.0^{+95.1}_{-95.1}\) \\
& empirical prior & \(-35.5^{+56.3}_{-61.4}\) & \(-27.9^{+31.5}_{-33.3}\) & \(-26.2^{+23.4}_{-24.4}\) \\
& Improv. & 39.4\% & 65.8\% & 74.9\% \\
\hline
\end{tabular}
\end{table}

\rep{With the weak $\bphi$ prior, the marginalized $\fnl$ constraints in \texttt{bfnl} mode show 68\% intervals from 90 to 100 in either direction. Because the weak prior has $\bphi$ priors symmetric to 0, the $\fnl$ posteriors are also symmetric about zero by construction. Consequently, their zero medians do not recover $\fnl^{\rm true}=\pm30$. Including the HOD-informed empirical $\bphi$ prior changes this behavior. The true value lies well within the central 68\% interval in each single-tracer fit, and is close to the medians. The full interval widths decrease by $38$--$79\%$ relative to the weak-prior baseline. The improvements are generally larger at higher redshift in these mocks as $\fnl$ signals are more sensitive in the higher galaxy bias case. }

%\clearpage
\subsection{Comparison with the Universality Relation}
\label{sec:prior vs universality}

\begin{table}[tb]
\centering
\caption{\rep{LRG3 marginalized \(\fnl\) constraints for the weak-prior \texttt{bfnl} reconstruction, the empirical-prior \texttt{bphi} fit, and the fixed-\(p=1\) universality model. The \(p=1\) column uses the existing \(k_{\max}=0.10\hmpcinv\) chains; their likelihood was replayed against the present data and covariance, but an independent convergence audit remains desirable.}}
\label{tab:bphi-prior-methods}
\resizebox{\columnwidth}{!}{%
\begin{tabular}{cccc}
\hline
\(\fnl^{\rm true}\) & \texttt{bfnl}, weak prior & \texttt{bphi}, empirical prior & \(p=1\) \\
\hline
0 & \(0.0^{+79.7}_{-79.7}\) & \(-1.6^{+22.3}_{-23.3}\) & \(-0.7^{+23.8}_{-23.3}\) \\
+30 & \(0.0^{+100.7}_{-100.7}\) & \(31.7^{+20.5}_{-21.4}\) & \(33.3^{+23.1}_{-22.2}\) \\
-30 & \(0.0^{+95.1}_{-95.1}\) & \(-26.2^{+23.4}_{-24.4}\) & \(-27.2^{+23.7}_{-23.4}\) \\
\hline
\end{tabular}}
\end{table}

After validating the pipeline, we compare different treatments of the PNG bias relation for LRG3, the sample with the tightest $\fnl$ constraint, as summarized in \cref{tab:bphi-prior-methods}. \rep{In these mocks, our HOD-informed empirical prior and the universality relation with $p=1$ give $\fnl$ constraints of comparable width. The universality result depends on a fixed relation between $\bphi$ and the galaxy bias, whereas the HOD-informed prior just need a distribution of $\bphi$ values.} Deviations from the assumed universality relation have been reported for galaxies and quasars \citep[e.g.,][]{Adame:2023nsx,Fondi:2026ilz}. Such deviations may become increasingly important for next-generation surveys such as MUST~\citep{Zhao:2024alp,cai2026}, Spec-S5~\citep{Spec-S5:2025uom}, and WST~\citep{WST:2024rai}. The HOD-informed prior therefore provides an alternative to inform $\bphi$ without undermining the precision of $\fnl$ measurements.

\subsection{Low-Redshift $\fnl$ Measurements}
\label{sec:volume}

\begin{table*}[tb]
\centering
\caption{\rep{Empirical-prior \texttt{bphi}-mode constraints for LRG3, LRG2--LRG3 and LRG1--LRG2--LRG3. Improvements are reductions in the full central 68\% interval width relative to LRG3 alone. The joint analyses assume independent per-bin response priors and a block-diagonal covariance.}}
\label{tab:bphi-prior-footprint}
\begin{tabular}{cccccc}
\hline
\(\fnl^{\rm true}\) & LRG3 & LRG2--LRG3 & Improv. & all LRGs & Improv. \\
\hline
0 & \(-1.6^{+22.3}_{-23.3}\) & \(-1.5^{+17.8}_{-18.2}\) & 21.3\% & \(-0.8^{+17.1}_{-17.7}\) & 23.6\% \\
+30 & \(31.7^{+20.5}_{-21.4}\) & \(31.6^{+16.6}_{-17.2}\) & 19.3\% & \(31.6^{+16.1}_{-16.2}\) & 22.8\% \\
-30 & \(-26.2^{+23.4}_{-24.4}\) & \(-26.6^{+18.7}_{-19.5}\) & 20.1\% & \(-26.7^{+17.8}_{-18.4}\) & 24.3\% \\
\hline
\end{tabular}
\end{table*}

As indicated by \cref{eq:PNG_Pgg}, tracers at higher redshift with larger galaxy bias can show a stronger scale-dependent PNG signal. Meanwhile, \cref{tab:veff} shows that the potential effective volume of LRG2 is similar to that of LRG3, and that LRG1 also covers a substantial volume. We therefore quantify the benefit of including lower-redshift LRGs by comparing LRG3 alone with the LRG2--LRG3 and three-sample joint fits in \cref{tab:bphi-prior-footprint}. \rep{Adding LRG2 reduces the full central 68\% interval width by approximately $19$--$21\%$ relative to LRG3 alone; including LRG1 and LRG2 reduces it by $23$--$24\%$ relative to LRG3 alone.} The true $\fnl$ value lies within the central 68\% interval of each fiducial three-sample fit. Note that these results depend on the adopted effective volumes and independent redshift-bin priors.

\subsection{Assembly bias}
\label{sec:assembly bias}
Because $\bphi$ responds to secondary halo properties beyond mass \citep{Lazeyras:2022koc,Fondi:2023egm,Fondi:2026ilz}, any $\bphi$ prior and the associated $\fnl$ inference pipeline should be tested against such secondary dependence.

We therefore test the robustness of our empirical $\bphi$ prior and $\fnl$ inference to one halo-environment assembly-bias prescription for LRG3. We consider three cases: (1) the standard case, without any assembly bias; (2) $P_{\rm AB}$, where only the data vector includes assembly bias; and (3) $P_{\rm AB}+{\rm prior}_{\rm AB}$, where both the data vector and the empirical $\bphi$ prior include it. In \cref{tab:bphi priors} and \cref{fig:bphi priors}, we already illustrated that this SDSS-type assembly bias\cite{Yuan:2021izi} did not shift much the $\bphi priors$. In \cref{tab:bphi-prior-assembly}, the marginalized $\fnl$ constraints are also similar across the three cases. 

\begin{table}[tb]
\centering
\caption{\rep{Effect of assembly bias on LRG3 marginalized \(\fnl\) constraints in \texttt{bphi} mode. \(P_{\rm AB}\) denotes the assembly-biased data vector, and \({\rm prior}_{\rm AB}\) the empirical prior measured with the same assembly-bias prescription.}}
\label{tab:bphi-prior-assembly}
\resizebox{\columnwidth}{!}{%
\begin{tabular}{cccc}
\hline
\(\fnl^{\rm true}\) & standard & \(P_{\rm AB}\) & \(P_{\rm AB}+{\rm prior}_{\rm AB}\) \\
\hline
0 & \(-1.5^{+22.6}_{-23.2}\) & \(0.0^{+21.0}_{-22.3}\) & \(-0.2^{+21.3}_{-21.7}\) \\
+30 & \(31.7^{+20.7}_{-21.6}\) & \(32.7^{+19.9}_{-20.3}\) & \(32.8^{+20.2}_{-20.3}\) \\
-30 & \(-26.5^{+23.2}_{-24.5}\) & \(-27.6^{+22.1}_{-24.0}\) & \(-28.6^{+22.5}_{-23.2}\) \\
\hline
\end{tabular}}
\end{table}

\begin{figure}[tb]
  \centering
  \includegraphics[width=\linewidth]{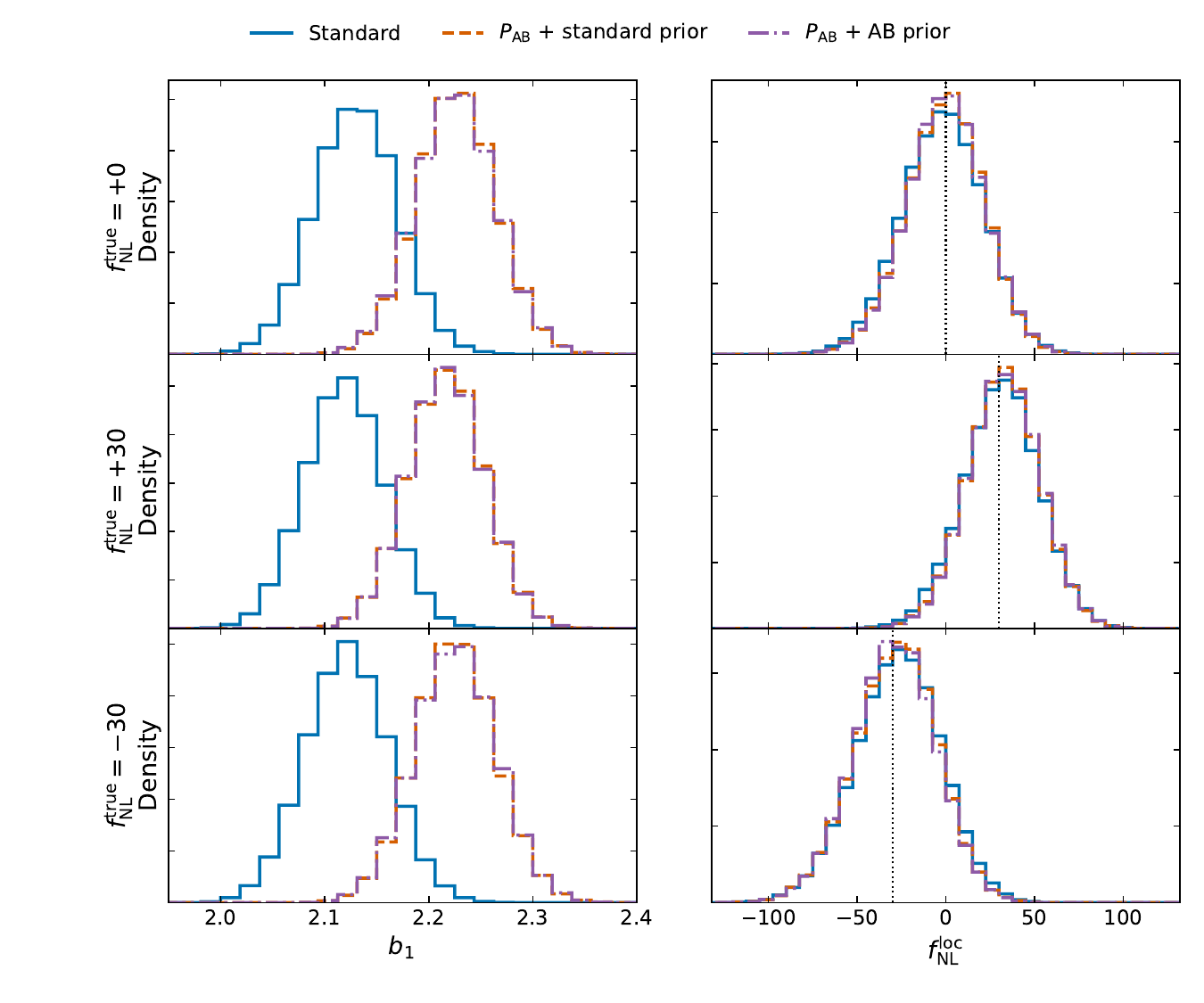}
  \caption{\rep{LRG3 marginal posteriors of \(b_1\) (left) and \(\fnl\) (right), for standard mocks and prior, assembly-biased mocks with the standard prior, and assembly-biased mocks with the assembly-biased prior. Rows correspond to \(\fnl^{\rm true}=0,+30,-30\) and dotted lines mark these \(\fnl^{\rm true}\) values. All fits use \texttt{bphi} mode with empirical priors. }}
  \label{fig:posterior assembly lrg3}
\end{figure}

The stability of the $\fnl$ measurements can be considered alongside the nuisance-parameter response in \cref{fig:posterior assembly lrg3}. The assembly-biased data vector shifts the marginal $b_1$ distribution upward by about $0.10$, compared with the  posterior standard deviation of $b_1$ about $0.04$, while the $\fnl$ shifts are small relative to their interval widths. Thus, the assembly bias has negligible influence on the empirical $\bphi$ priors construction and the final $\fnl$ constraints. We should emphasise that this is a targeted test of one prescription measured from SDSS LRGs, rather than a general demonstration of immunity to assembly bias for $\fnl$ measurements.

\section{Summary and Outlook}
\label{sec:conclusion}

Robust and precise constraints on the local PNG amplitude $\fnl$ from galaxy clustering require external knowledge of $\bphi$. We have proposed a third route to it. Rather than calibrating $\bphi$ with galaxy-formation simulations or deriving it from the observed $dn/dz$, we map the small-scale clustering of the dark matter tracers onto empirical $\bphi$ priors through observation-calibrated galaxy--halo connections. \rep{Considering realistic LRG samples, we used the DESI LRG HOD posteriors reported by \citet{Yuan:2023ezi}, populate model galaxies in pairs of standard and low-$\sigma_8$ \abacus~simulations, and measure $\bphi$ from the separate-universe method with the pairs of galaxy mocks. We use the resulting 100 $\bphi$ values to construct an empirical prior for each redshift bin. For LRG1 and LRG2 representing LRGs at $0.4<z<0.6$ and $0.6<z<0.8$, we broaden the HOD posteriors by a factor of three. For LRG3 for LRGs at $0.8<z<1.1$, we use the raw posterior given its loose parameter constraints. This is to make the galaxy--halo relation realistic and generic to any survey.}

On the data-vector $\abacuspng$ mocks with $\fnl^{\rm true}=0,\pm30$, the central 68\% intervals from all single-tracer fits with the HOD-informed priors contain the true $\fnl$ values. \rep{Their interval widths are 38--79\% smaller than those obtained with our baseline weak $\bphi$ prior, with the largest gains at high redshift. For LRG3, the uninflated HOD chains provided empirical $\bphi$ priors that yield comparable $\fnl$ constraint with that from the universality relation with $p=1$. However, we do not assume a fixed relation between $\bphi$ and the galaxy bias. Fitting LRG1--LRG2--LRG3 jointly reduces the interval width by 23--24\% relative to LRG3 alone, and the major contribution is from LRG2 with a higher galaxy bias and larger volume.} The LRG3 $\fnl$ constraints are also stable under the SDSS-type assembly-bias prescription tested here.

%\mn{Check this!}\jy{Have you finished checking the limitation section? 12 Nov 2026}
Several limitations bound these results and this pilot study. The analysis is restricted to LRGs and to one mock realization per PNG amplitude. Extending the approach to ELGs and quasars is not automatic, as the quasar HOD is less well understood and the low galaxy bias and complicated systematics control of ELGs making it less promising in PNG measurements. \rep{The data vector and covariances are constructed by cubic mocks with hypothetical effective volumes rather than realistic mocks including survey geometry effect and observational systematics. Before implementing this method to real DESI data, we need to elaborate on the cross-correlation between the small-scale HOD information and the large-scale $\fnl$ constraints. Therefore, the constraints and quoted improvements are idealized and may change when applied to real DESI data. The empirical priors are constructed from only 100 accepted HOD draws per sample. Although our sampling retains correlations within each HOD posterior, inflating HOD posteriors extends the width of $\bphi$ distributions, and thus weakens the $\fnl$ constraints, especially for low-redshift samples.} The assembly-bias test is limited to LRG3 and one fixed prescription from SDSS.

None of the three approaches to obtaining $\bphi$ mentioned above is decisive on its own, and they fail differently: simulation calibration inherits the galaxy-formation model, the $dn/dz$ route rely on the accurate modeling of the survey selection functions, and ours depends on the HOD constraints. Their disagreement is therefore informative, and combining them will provide important cross-validations. As small-scale clustering measurements and our understanding of galaxy formation sharpen, the informative $\bphi$ priors will sharpen with them, improving the prospects of using the large-scale imprint of local PNG to probe the physics of inflation with Stage-V surveys such as MUST~\citep{Zhao:2024alp,cai2026}, Spec-S5~\citep{Spec-S5:2025uom}, and WST~\citep{WST:2024rai}.

\medskip

\begin{acknowledgments}
JXY thanks Antoine Rocher and Hanyu Zhang for helpful discussions.
NMN thanks the organizers of the Croucher Advanced Study Institute ``Primordial Non-Gaussianities'' for hospitality.
This work was performed using the \code{NERSC} at the LBNL and the \code{idark} cluster at KIPMU.
JXY acknowledges support from the National Natural Science Foundation of China (NSFC) grant No.12573002 and the Japan Society for the Promotion of Science (JSPS) KAKENHI Grant Number 26K17134.
NMN acknowledges support from the Japan Foundation for Promotion of Astronomy Research Grant and the JSPS KAKENHI Grant Number 25K23373 and 26H00404. 
This work was supported by World Premier International Research Center Initiative (WPI Initiative), MEXT, Japan.

\rep{We used OpenAI Codex, an AI coding assistant based on GPT-5 and GPT-6, to assist with code development, diagnostic checks, and the preparation of figures and text. The authors directed all tasks, reviewed and verified the outputs, and take responsibility for the scientific analysis and conclusions.}
\end{acknowledgments}

\clearpage
\onecolumngrid
\appendix
\crefalias{section}{appendix}
\crefalias{subsection}{appendix}

\section{Marginal posterior of $\fnl$}
\label{app:margpost}

\rep{In the fiducial \texttt{bphi} analysis, we sample $\fnl$ directly,
together with $\bphi$ values drawn from the empirical prior in each
redshift bin. We obtain the marginalized $\fnl$ posterior by
marginalizing over $\bphi$ and the other model parameters. In the
joint fit, $\fnl$ is shared across redshift bins, while the empirical
$\bphi$ priors are assumed to be independent.}

\rep{For the weak-prior \texttt{bfnl} analysis, the sampled parameter is
instead the product $B=\bphi\fnl$. To calculate the marginalized
$\fnl$ posterior, we use the additional flat priors on $\bphi$ and
$\fnl$ described in \cref{sec:prior}. For each retained sample $B_j$ and a proposed value $f=\fnl$, the corresponding value of $\bphi$ is $B_j/f$. If this value lies
outside the assumed range $[-10,10]$, the combination is discarded.
Otherwise, the sample contributes to the posterior density at $f$ with a weight proportional to $1/|f|$:
\begin{equation}
 W_j(f)\propto
 \begin{cases}
  1/|f|, & \text{if } -10\leq B_j/f\leq10,\\
  0, & \text{otherwise},
 \end{cases}
 \qquad 0<|f|\leq300.
 \label{eq:posthoc_lnW}
\end{equation}
The factor $1/|f|$ accounts for the change of variables from $\bphi$ to $B$.  We add these weights at each $f$ and normalize the resulting distribution so that its total area is one. For a joint fit, we apply the $\bphi$-range condition and a factor of $1/|f|$ for each redshift bin.}

\rep{Because the weak $\bphi$ prior gives equal probability to positive
and negative values, the resulting $\fnl$ posterior is symmetric
about zero, regardless of the true $\fnl$ value. Its width depends
on the assumed $\bphi$ range, so this is not a prior-free constraint
on $\fnl$.}

\section{Robustness to the choice of $\kmax$}
\label{app:kmax_robustness}

\begin{figure}[tb]
  \centering
  \includegraphics[width=\linewidth]{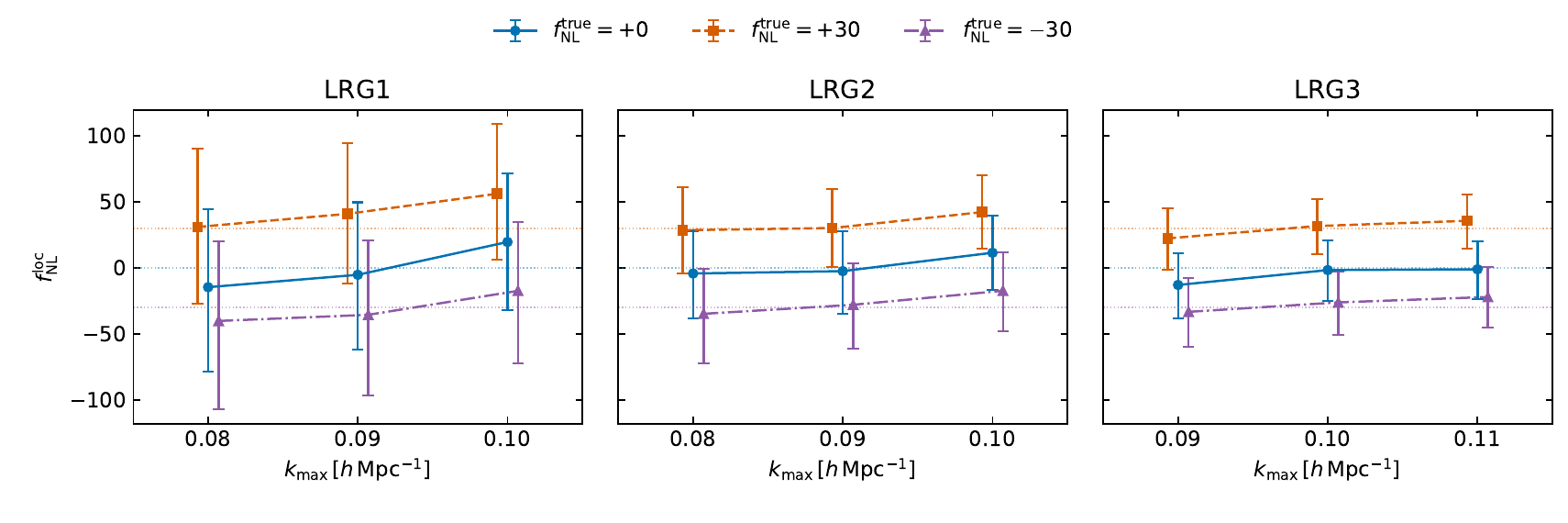}
  \caption{\rep{Dependence of the marginalized $\fnl$ constraints on the
maximum fitting wavenumber, using empirical $\bphi$ priors. The columns
show LRG1, LRG2, and LRG3. Points show posterior medians and error bars
indicate the central 68\% credible intervals. Colors and markers
distinguish the mocks with $\fnl^{\rm true}=0,+30,-30$, with the
corresponding true values shown by dotted horizontal lines. Points are
shifted horizontally for visualization.}}
  \label{fig:posterior kmax}
\end{figure}

\rep{We compare single-tracer empirical-prior fits for \(\fnl^{\rm true}=0,\,+30,\,-30\) mocks in \cref{tab:fnl-kmax} and \cref{fig:posterior kmax}, keeping \(k_{\rm min}=0.006\hmpcinv\). Though DESI used $\kmax=0.08\hmpcinv$ for LRG3, we would like to explore the largest possible $\kmax$ values for all our samples. For LRG1 and LRG2 we show \(\kmax=0.08,0.09,0.10\hmpcinv\); for LRG3 \(0.09,0.1,0.11\hmpcinv\). The true $\fnl$ value are supposed to lie within the center of the 68\% intervals, and the widths become wider as \(\kmax\)\ decreases. With \(k_{\rm max}=0.10\hmpcinv\), the medians of the posteriors for LRG1 and LRG2 both deviate from the true value, while for \(k_{\rm max}=0.08\hmpcinv\) the medians are close to the true values but the constraints are worse than those of \(k_{\rm max}=0.09\hmpcinv\). For the same reason, we chose $\kmax=0.10\hmpcinv$ for LRG3. Therefore, we will use \(k_{\rm max}=0.09\hmpcinv\) for LRG1 and LRG2, and $\kmax=0.1\hmpcinv$ for LRG3 for all the fittings. The larger $\kmax$ of LRG3 is because $\kmax=0.10\hmpcinv$ still belongs to linear regime at this higher redshift bin.  }
\begin{table}[htbp]
\centering
\caption{\rep{Marginalized $\fnl$ constraints for different $\kmax$, using empirical $\bphi$ priors. Entries give the posterior
median and central 68\% credible interval.}}
\label{tab:fnl-kmax}
\begin{tabular}{ccccc}
\hline
$\fnl^{\rm true}$ & $k_{\max}\,[h\,\mathrm{Mpc}^{-1}]$
& LRG1 & LRG2 & LRG3 \\
\hline
\multirow{4}{*}{$0$}
& 0.08 & $-14.6^{+59.4}_{-64.2}$ & $-4.2^{+32.4}_{-34.2}$ & -- \\
& 0.09 & \bm{$-5.2^{+54.9}_{-56.9}$}  & \bm{$-2.4^{+30.2}_{-32.1}$} & $-12.9^{+24.1}_{-25.4}$ \\
& 0.10 & $19.6^{+51.7}_{-51.8}$  & $11.5^{+28.4}_{-27.9}$ & \bm{$-1.6^{+22.3}_{-23.3}$ }\\
& 0.11 & -- & -- & $-1.0^{+21.3}_{-22.4}$ \\
\hline
\multirow{4}{*}{$+30$}
& 0.08 & $31.0^{+59.4}_{-58.4}$ & $28.5^{+32.7}_{-32.7}$ & -- \\
& 0.09 & \bm{$40.9^{+53.7}_{-52.6}$} & \bm{$30.2^{+29.5}_{-29.2}$} & $22.4^{+22.9}_{-23.5}$ \\
& 0.10 & $56.2^{+52.8}_{-49.7}$ & $42.2^{+27.8}_{-27.6}$ & \bm{$31.7^{+20.5}_{-21.4}$} \\
& 0.11 & -- & -- & $35.7^{+19.8}_{-20.9}$ \\
\hline
\multirow{4}{*}{$-30$}
& 0.08 & $-40.2^{+60.6}_{-67.0}$ & $-34.8^{+34.0}_{-37.3}$ & -- \\
& 0.09 & \bm{$-35.5^{+56.3}_{-61.4}$} & \bm{$-27.9^{+31.5}_{-33.3}$} & $-33.3^{+25.7}_{-26.4}$ \\
& 0.10 & $-17.4^{+52.2}_{-54.7}$ & $-17.3^{+29.2}_{-30.7}$ & \bm{$-26.2^{+23.4}_{-24.4}$} \\
& 0.11 & -- & -- & $-22.0^{+22.9}_{-23.3}$ \\
\hline
\end{tabular}
\end{table}

\section{The Accuracy of the Approximate Gaussian $\bphi$ Priors}
\label{app:gaussian-bphi}
%\mn{Shouldn't this appendix be highlighted as well?}\jy{Done}
\rep{
In this section, we assess the effect of approximating the $\bphi$ distributions, which are derived from the HOD posteriors using the separate-universe method, by Gaussian distributions. We compare the $\fnl$ posteriors inferred using the fiducial $\bphi$ priors directly against those inferred with Gaussian $\bphi$ priors whose means and variances match the corresponding fiducial ones.}

\rep{We use the broadened $s=3$ HOD ensembles for LRG1 and LRG2 and the unbroadened $s=1$ ensemble for LRG3. For each prior choice, we sample the common $\fnl$ directly in \texttt{bphi} mode, using the same mock data, covariance and tracer-dependent scale cuts $\kmax=(0.09,0.09,0.10)\hmpcinv$. We consider each LRG sample separately and the joint three-sample fit for mocks with $\fnl^{\rm true}=0,+30,-30$.}

\begin{figure*}[tb]
  \centering
  \includegraphics[width=\linewidth]{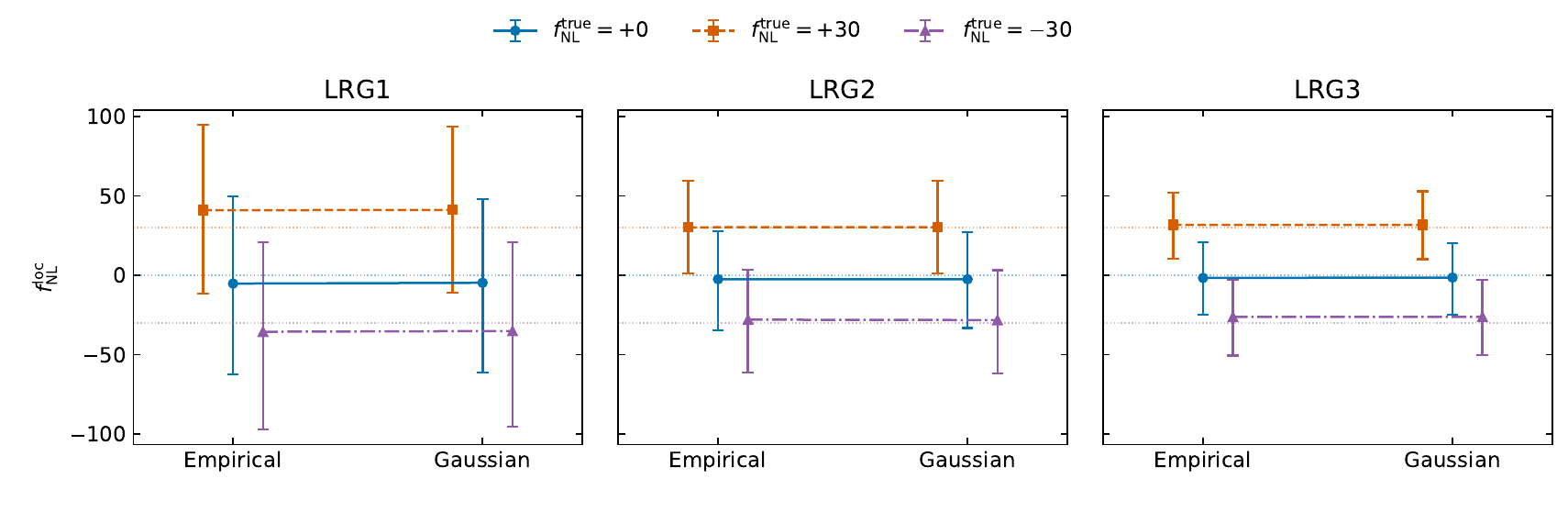}
  \caption{\rep{Similar to \cref{fig:posterior kmax} but for comparison of marginalized $\fnl$ constraints obtained using
empirical and Gaussian $\bphi$ priors for LRG1, LRG2, and LRG3.}}
  \label{fig:gaussian-empirical-fnl}
\end{figure*}

\rep{As shown in \cref{fig:gaussian-empirical-fnl}, the empirical and Gaussian priors produce similar marginalized $\fnl$ constraints in these tests: the full central 68\% credible interval widths differ by at most $3.4\%$, and the posterior medians differ by at most $0.5$. Thus, although a Gaussian need not reproduce the detailed shape of a measured $\bphi$ distribution, it provides a similar $\fnl$ constraint to the empirical priors.}

\bibliographystyle{apsrev4-2}
\bibliography{references}

\end{document}